\documentclass[%
 reprint,
 amsmath,amssymb,
 aps,
 prl,
]{revtex4-2}

\usepackage{graphicx}
\usepackage{dcolumn}
\usepackage{bm}

\begin{document}

\preprint{APS/123-QED}

\title{A General Relation Between the Largest Nucleus and All Nuclei Distributions for Free Energy Calculations}

\author{Jo\"el Puibasset}
 \email{puibasset@cnrs-orleans.fr}
\affiliation{%
 ICMN, CNRS, Universit\'e d'Orl\'eans, 1b rue de la F\'erollerie, CS 40059, 45071 Orl\'eans cedex 02, France}%


\date{\today}

\begin{abstract}
Prediction of nucleation rates in first order phase transitions requires the knowledge of the barrier associated to the free energy profile $W$.
Molecular simulations offer a direct route through $W = -kT \ln p_a$, where $k$ is Boltzmann's constant, $T$ is temperature, and $p_a$ the probability distribution of the size of \emph{any} nucleus.
But in practice, the extremely scarce spontaneous occurrence of large nuclei impedes the full determination of $p_a$, and a numerical bias must be introduced, \textit{e.g.} on the size of the largest nucleus in the system, leading to the probability size distribution of the \emph{largest} nucleus $p_l$.
Although $p_l$ is known to be system size dependent, unlike $p_a$, it has been extensively used as an approximation for $p_a$. 
This paper proposes an exact relation between $p_a$ and $p_l$, which cures this approximation and allows an exact calculation of free energy barriers from biased simulations.     
\end{abstract}

\maketitle


First order phase transitions generally occur in the metastable phase by nucleation of the stable one via an activated process.
According to the classical nucleation theory, the new stable phase has to grow spontaneously from the parent metastable one until it reaches a critical size beyond which the process is irreversible \cite{RN548,RN512,RN2920}. 
Since only one successful event is required, the transition may occur for barriers as large as several tens of $kT$, where $k$ is Boltzmann's constant and $T$ is temperature.

Atomistic simulations are essential to understand the first stages of nucleation at the molecular level, generally inaccessible to experiments. 
They are also useful to bridge the gap between theory and experiments and understand discrepancies.
From a statistical point of view, the free energy profile $W(s)$ of the growing nucleus of size $s$ relates to the equilibrium probability $p_a(s)$ that the size of \emph{any} nucleus is $s$ through \cite{RN2859,RN2913}:
\begin{equation}
 W(s) = -kT \ln p_a(s) \label{eq_boltz}.
\end{equation}

In principle, $p_a$ is measurable in equilibrium molecular simulations. 
However, except for high temperature or superheat, the barrier is too high to be spontaneously overcome in simulations (due to limited system size and running time), and, in practice, $p_a$ cannot be sampled 
except for the smallest nuclei.
It is therefore necessary to perform biased simulations where the nucleus size is monitored all the way from the bottom to the top of the barrier. In a seminal study of homogeneous crystal nucleation of soft repulsive spheres by molecular simulation \cite{RN2901,RN2909}, it was proposed to monitor the size of the largest nucleus as a local order parameter 
by means of the umbrella scheme of Torrie and Valleau \cite{RN2885}. 
This method leads to the probability distribution of the \emph{largest} nucleus, denoted $p_l (s)$. 
The authors argue that the obtained free energy profile $-kT \ln p_l (s)$ can be identified to that of the growing nucleus $W(s)$ (\textit{i.e.} $p_a \propto p_l$) since there is essentially only one large nucleus in the system \cite{RN2901,RN2909}.

This method has been a breakthrough for exploring the nucleation barrier of various systems \cite{RN1956, RN2911, RN2915, RN2889}. However, it was rapidly recognized that $p_l$ is actually \emph{not} suitable for free energy calculations in the first stages of nucleation and must be replaced by the exact $p_a$ \cite{RN2905, RN2893, RN2895, RN2907, RN2903, RN2847, RN2801}.
Hopefully, $p_a(s)$ is accessible to molecular simulations for small $s$, \textit{e.g.} $s \le s_0$. For $s \ge s_0$, one still uses $p_l(s)$, which is multiplied by a constant to ensure continuity at $s_0$.  
This approach, denoted hereafter the $s_0$-method, has however severe limitations pointed out very recently by Goswami and coworkers \cite{RN2899} : (i) the critical nucleus may fall in the region around $s_0$, which should be avoided to prevent barrier dependence on the exact value chosen for $s_0$, (ii) the simulation box should be chosen large enough to avoid cross-talking of nuclei through periodic boundary conditions, but (iii) not too large to avoid multiple nuclei and ensure $p_l \propto p_a$. 
Considering the fact that there are multiple situations where such constrains cannot be fulfilled, it has become mandatory to derive an exact relationship between $p_a$ and $p_l$.

To illustrate our theoretical development, we consider the special case of cavitation in liquids, but it applies to condensation, crystallization, etc. 
The system is made of atoms interacting via the truncated and shifted Lennard-Jones potential with a cutoff radius equal to 2.5 atomic diameters, in thermodynamic conditions identical to those considered in Ref.~[\onlinecite{RN2797}], \textit{i.e.} at constant reduced temperature $kT/\epsilon=0.855$ and constant reduced pressure $p\sigma^3/\epsilon=0.026$, where $\epsilon$ and $\sigma$ are the Lennard-Jones parameters (Monte Carlo simulations in the isothermal-isobaric $NPT$ ensemble). 
In these conditions the liquid is metastable (superheated). Three system sizes are considered: $N =$ 442, 3375 and 8000 atoms.

The bubbles appearing in the system are detected and characterized according to the M-method \cite{RN1184,RN2847}: 
(i) For each atom, if the number of neighbors closer than $1.6\sigma$ is larger or equal to 6 it is labeled as liquid-like, and vapor-like otherwise.
(ii) The simulation box is partitioned into cubic voxels of edge approximately $0.5\sigma$, the exact dimension $l$ being adjusted during the simulation run to the fluctuating box size. 
(iii) Voxel centers closer than $1.6\sigma$ from a liquid-like atom are marked as liquid; the remaining are marked as vapor.
(iv) A cluster analysis is performed on the vapor voxels, with a criterion of spatial connection through face sharing. This procedure allows to characterize the bubbles appearing in the system in each configuration, \textit{i.e.} their number and sizes defined as the total number of voxels in each cluster. 
Of course, the exact values obtained for the bubble sizes depend on the criteria used to characterize them. 
It has however been shown that the different methods that have been proposed yield similar results \cite{RN2849, RN2847}.
It is thus meaningful to calculate the two histograms associated to all bubble sizes and to the largest one, and identify them with $p_a$ and $p_l$ respectively.  
Formally, these discrete histograms, denoted with a tilde to distinguish them from the continuous densities, are defined as follows: 
\begin{eqnarray}
 \widetilde{p_a}(i) =  \int_{i \nu}^{(i+1) \nu} p_a (s)ds  \label{eq_disc1}
\\
 \widetilde{p_l}(i) =  \int_{i \nu}^{(i+1) \nu} p_l (s)ds \label{eq_disc2}
\end{eqnarray}
where $i$ is an integer $\geq 0$, $\nu$ is the volume of one voxel, and $p_a$ and $p_l$ are the density distributions normalized to 1, so that the histograms are also normalized to 1.
Note that, by construction, these histograms depend on $\nu$. 
By definition, $\widetilde{p_l}(0)$ is the probability that the largest bubble is less than one voxel, \textit{i.e.} the
probability that the molecular configuration contains no bubble detectable with the discretized procedure. 
$\widetilde{p_a}(0)$ corresponds to the probability that the volume of any bubble is less than one voxel; this quantity is \emph{not} accessible by numerical calculations since these bubbles are not detectable with the discretized approach. Therefore, in practice, $\widetilde{p_a}(i)$ can be calculated only for $i \geq 1$, and its renormalization to 1 gives
$\widetilde{p_a}(i \geq 1)/\alpha$ with $\alpha = \sum_{i=1}^\infty \widetilde{p_a}(i) = \int_{\nu}^{\infty} p_a(s)ds$. 

\begin{figure}[b]
\includegraphics{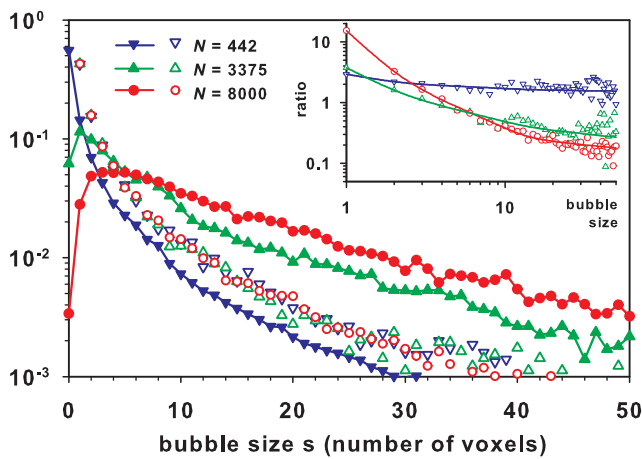}
\caption{\label{fig_p_and_pL} Normalized histograms of bubble sizes for the three systems $N=442, 3375$ and $8000$. Empty symbols: $\widetilde{p_a}(i \geq 1)/\alpha$ (all bubbles); filled symbols and lines: $\widetilde{p_l}(i \geq 0)$ (largest bubble). Inset: ratio between the two histograms $\widetilde{p_a}/\alpha \widetilde{p_l}$; lines are guides to the eye. 
}
\end{figure}
Figure~\ref{fig_p_and_pL} gives the normalized histogram distribution of all bubbles $\widetilde{p_a}(i \geq 1)/\alpha$ and the histogram distribution of the largest bubble $\widetilde{p_l}(i \geq 0)$ during the simulation run for the three system sizes. As can be seen:

(i) $\widetilde{p_a}$ and $\widetilde{p_l}$ have similar behavior for the smallest system ($N=442$), where it is highly improbable to have more than one bubble, but significantly differ for larger systems, in particular for small bubbles. 

(ii) $\widetilde{p_a}$ always decreases, while $\widetilde{p_l}$ may exhibit a maximum for a large enough system, because the number of simultaneous bubbles is larger, and therefore it is less probable that the largest one be small \cite{RN2895}. 

(iii) $\widetilde{p_a}$ and $\widetilde{p_l}$ tend to become proportional at large bubble sizes (their ratio tend to be constant, see inset in Fig.~\ref{fig_p_and_pL}). 
However, the convergence is not reached before the degradation of precision (data dispersion).
Therefore, the connection of the two histograms with the $s_0$-method necessarily introduces a non-controlled error whatever the choice for $s_0$ (rediscussed at the end). 

(iv) Uncertainties on $\widetilde{p_a}$ grow rapidly with the nucleus size due to the energetic cost of spontaneous nucleation, while $\widetilde{p_l}$ is very accurate thanks to biased sampling
, explaining its interest.  

(v) $\widetilde{p_l}$ depends on system size: it is not an intrinsic property of the fluid, reducing its physical interest. 

Points (iv) and (v) summarize the dilemma: $\widetilde{p_l}$ is accessible to simulations up to the critical size, while $\widetilde{p_a}$ is the required quantity enterring Eq.~(\ref{eq_boltz}). 
We now derive a quantitative relationship between $\widetilde{p_a}$ and $\widetilde{p_l}$ to bridge the gap between simulation results and nucleation barriers.

If exactly one nucleus occurs in the system, $p_l=p_a$.
If exactly two independent nuclei occur, the probability that the largest has a size $s$ is the probability that one has the size $s$, given by $p_a(s)$, while the other has a size less or equal to $s$, given by the cumulative distribution function (CDF) $P_a(s) = \int_0^s p_a(s) ds$. This translates into  $p_l(s) = 2 p_a(s)P_a(s)$.
If the number of nuclei in the system is exactly $m$, this generalizes to $p_l(s) = m p_a(s)P_a(s)^{m-1}$.
By integration, 
one finds that the CDF associated to $p_l$ verifies $P_l (s) = P_a(s)^m$, up to a constant which vanishes since $p_a$ and $p_l$ are normalized. 
The relation is known as the distribution of the maximum of $m$ independent and identically distributed variables.
The case $m=0$ deserves special attention: it corresponds to the situation where no nucleus occurs, and one gets $P_l (s>0)=1$, associated to a Dirac distribution for $p_l$.

Let us now suppose that the number of nuclei $m$ in the system follows a normalized distribution, denoted $\phi(m)$. The CDF $P_l (s)$ is now expected to be the weighted sum:
\begin{equation}
 P_l(s) = \sum_{m=0}^\infty \phi(m) P_a(s)^m   \label{eq_Pgeneral}
\end{equation}
where the sum extends to all possible values for the number of nuclei in the system. For $s \rightarrow \infty$, $P_a(s)\rightarrow 1$, and $\phi$ being normalized, one recovers the expected result $P_l (\infty)=1$. For $s\rightarrow 0$, all non-zero powers $P_a(s)^m \rightarrow 0$, and $P_l (0) = \phi(0)$, which corresponds as expected to the probability to have no nucleus in the system. 

What can be said about $\phi$? Nucleation being a rare process, it is unlikely that two nuclei interact 
and they can safely be considered as independent. The number of nuclei at a given time then follows a Poisson distribution:
\begin{equation}
 \phi(m) = e^{-\lambda_0}\frac{\lambda_0^m}{m!}  \label{eq_Poisson}
\end{equation}
where $\lambda_0$ is the average number of nuclei in the system (proportional to the system size). Note that the simple power law $\phi(m)\propto\lambda_0^m$ previously proposed \cite{RN2911, RN2913} does not take into account the fact that the nuclei appear in the same volume.
\begin{figure}[b]
\includegraphics{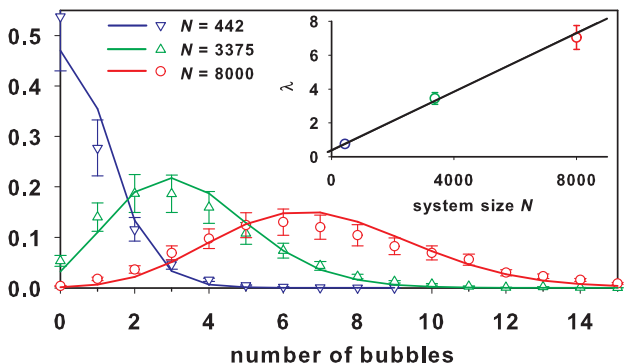}
\caption{\label{fig_poisson} Histograms of the number of bubbles in the system as given by molecular simulations (symbols) and the corresponding Poisson distribution fits (lines, Eq.~(\ref{eq_Poisson})) for the three system sizes $N = 442, 3375$ and 8000. Inset: evolution of the Poisson parameter $\lambda$ (average number of bubbles, see text below Eq.~(\ref{eq_discPPoisson})) with the system size $N$.}
\end{figure}
Figure~\ref{fig_poisson} shows the numerical results for $\phi$ for the three system sizes. They are well fitted with Poisson distributions, and one verifies that the corresponding Poisson parameter is proportional to the system size (inset).
Using Eqs.~(\ref{eq_Pgeneral}) and (\ref{eq_Poisson}) one gets:
\begin{equation}
 P_l(s) = e^{-\lambda_0 (1-P_a(s) )} . \label{eq_PPoisson}
\end{equation}
Inversion of Eq.~(\ref{eq_PPoisson}) followed by derivation gives the desired relationship between the measured $p_l$ (and its CDF $P_l$) and the required $p_a$ enterring Eq.~(\ref{eq_boltz}):
\begin{equation}
 p_a(s) = \frac{p_l(s)}{\lambda_0 P_l(s)} .
\end{equation}
For $s \rightarrow \infty$, $P_l(s)\rightarrow 1$ : therefore $p_a$ and $p_l$ tend to become proportional, which justifies the generally admitted relation $p_a \propto p_l$ \cite{RN2901,RN2909}. 
Conversely, in the first stages of nucleation where $P_l$ is small, the two distributions clearly depart significantly. 

For numerical applications, one should consider the discretized histograms $\widetilde{p_a}$ and $\widetilde{p_l}$ instead of $p_a$ and $p_l$. Using Eqs.~(\ref{eq_disc1}) and (\ref{eq_disc2}), the CDF for $\widetilde{p_a}$ and $\widetilde{p_l}$ write 
\begin{eqnarray}
 \widetilde{P_a}(n) = \sum_{i=0}^n \widetilde{p_a}(i) = P_a((n+1)\nu)    \label{eq_discsum}
\\
 \widetilde{P_l}(n) = \sum_{i=0}^n \widetilde{p_l}(i) = P_l((n+1)\nu)    \label{eq_discsumL}
\end{eqnarray}
and thus the discretized version of Eq.~(\ref{eq_PPoisson}) writes: 
\begin{equation}
 \widetilde{P_l}(n) = e^{-\lambda_0 (1-\widetilde{P_a}(n))}  \text{ for } n \geq 0 . \label{eq_discPPoisson}
\end{equation}
For $n=0$, one gets $\widetilde{P_l}(0)=e^{-\alpha \lambda_0}$ with $\alpha = \sum_{i=1}^\infty \widetilde{p_a}(i)$. This is consistent with the interpretation that $\widetilde{p_l}(0) = \widetilde{P_l}(0)$ is the probability that a configuration contains no numerically detectable nucleus (on the discrete grid) thanks to the following argument: introducing a nucleus detection threshold $\nu$ (voxel) transforms the initial Poisson distribution Eq.~(\ref{eq_Poisson}) into a new Poisson distribution with parameter $\lambda = \alpha \lambda_0$, because the probability to observe $m$ nuclei larger than $\nu$ requires that there is at least $n=m$ nuclei in the system with probability $\phi(n)$, among which $m$ have the probability $\alpha$
to be larger than $\nu$. 

Combining Eqs.~(\ref{eq_discsum}) and (\ref{eq_discPPoisson}) gives :
\begin{equation}
 \lambda_0 \widetilde{p_a}(n) = \ln \widetilde{P_l}(n) - \ln \widetilde{P_l}(n-1)  \text{ for } n \geq 1 
\end{equation}
and for $n=0$, $\lambda_0 \widetilde{p_a}(0) = \lambda_0 + \ln \widetilde{P_l}(0)$ which is equivalent to the requirement that $\widetilde{p_a}$ is normalized to unity, and thus brings no new information to determine the unknown constant $\lambda_0$. 
However, as already mentioned, the first term $\widetilde{p_a} (0)$ is \emph{not} measurable (below the detection threshold of one voxel) and has to be discarded; the corresponding normalized histogram is $\widetilde{p_a} (i \geq 1)/\alpha$. Using the measurable average number of nuclei above the threshold $\lambda = \alpha \lambda_0 = -\ln \widetilde{p_l}(0)$ one finally gets :
\begin{equation}
 \widetilde{p_a}(n)/\alpha = \frac{1}{\lambda} \left\{ \ln \widetilde{P_l}(n) - \ln \widetilde{P_l}(n-1) \right\} \text{ for } n \geq 1  \label{eq_discp}
\end{equation}

This gives the \emph{numerical algorithm} to transform the measured $\widetilde{p_l}$ into $W(s)$: 
(i) establish the histogram $\widetilde{p_l}(i)$ for $i \geq 0$ using biased methods to improve statistics up to the critical nucleus size, 
(ii) deduce the average number of nuclei $\lambda = -\ln \widetilde{p_l}(0)$ and the discrete CDF $\widetilde{P_l}(n) = \sum_{i=0}^n \widetilde{p_l}(i)$, 
(iii) calculate $\widetilde{p_a}(n)/\alpha$ for $n \geq 1$ using Eq.~(\ref{eq_discp}), and,
(iv) calculate $W(s)$ using Eq.~(\ref{eq_boltz}), the irrelevant constant being fixed by choosing $W(0)$. 
Note that the algorithm does not require the explicit determination of $\phi$ and is therefore straightforward.

\begin{figure}[t]
\includegraphics{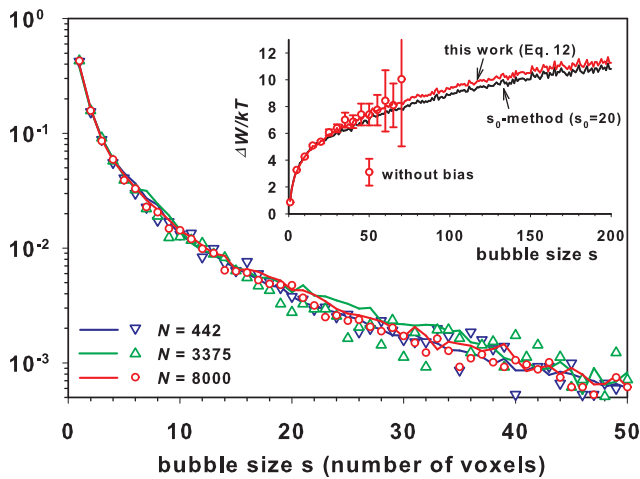}
\caption{\label{fig_p_and_p_from_pL} Histograms of all bubbles $\widetilde{p_a}(n \geq 1)/\alpha$ as given by Eq.~(\ref{eq_discp}) (lines) or given directly by simulations without bias (symbols, reported from Fig.~\ref{fig_p_and_pL}) for the three system sizes $N$ = 442, 3375 and 8000. Inset: the corresponding reduced free energies $\Delta W/kT$ given by Eq.~(\ref{eq_boltz}) for $N$=8000; for comparison, the result of the $s_0$-method is also shown (black line).}
\end{figure}

Figure~\ref{fig_p_and_p_from_pL} gives the results for the three system sizes: as can be seen, the distributions given by Eq.~(\ref{eq_discp}) (lines) perfectly superimpose to the direct calculation of $\widetilde{p_a}$ without bias (symbols) when both data are available. The excellent agreement down to the smallest possible nucleus size of one voxel proves the accuracy of this new method.  
The inset gives the corresponding free energy variations $\Delta W/kT$. For comparison, the result of the $s_0$-method is also shown: the small but visible disagreement reveals the error introduced by the connection between $p_l$ and $p_a$ in the $s_0$-method which results in a constant shift.

This work opens new perspectives in the calculation of nucleation barriers from molecular simulations with a triple advantage. (i) It gives a general procedure that applies for any nucleation phenomenon (cavitation, condensation, crystallization, etc.), (ii) the algorithm gives the exact distribution of all nuclei $p_a$ entering Eq.~(\ref{eq_boltz}) for any nucleus size, and (iii) the algorithm is simple and does not rely on any approximate adjustment procedure.

\begin{acknowledgments}
The author acknowledges fruitful discussions with P.E. Wolf, E. Rolley and P. Porion, and the financial support of Agence Nationale de la Recherche through the project CavConf, ANR-17-CE30-0002.
\end{acknowledgments}

\bibliography{article}

\end{document}